\pdfoutput=1
\documentclass[10pt,letterpaper]{article}
\usepackage[margin=1in]{geometry}
\usepackage[T1]{fontenc}
\usepackage{microtype}
\usepackage{float}
\usepackage{amsmath,amssymb,mathtools}
\usepackage{graphicx}
\usepackage{booktabs}
\usepackage{siunitx}
\usepackage{caption}
\usepackage{url}     
\usepackage{titling}
\usepackage[round,authoryear]{natbib}
\usepackage{titlesec}
\title{%
\textbf{CourtGuard: A Local, Multiagent Prompt Injection Classifier}\\[0.2em]
}

\author{%
\textbf{Isaac Wu} \\
\textit{Research Fellow} \\
Non-Trivial Ventures \\
\and
\textbf{Michael Maslowski} \\
\textit{Facilitator} \\
Non-Trivial Ventures \\
}

\begin{document}
\maketitle

\begin{abstract}
\noindent
As large language models (LLMs) become integrated into various sensitive applications, prompt injection, the use of prompting to induce harmful behaviors from LLMs, poses an ever increasing risk. Prompt injection attacks can cause LLMs to leak sensitive data, spread misinformation, and exhibit harmful behaviors. To defend against these attacks, we propose CourtGuard, a locally-runnable, multiagent prompt injection classifier. In it, prompts are evaluated in a court-like multiagent LLM system, where a ``defense attorney'' model argues the prompt is benign, a ``prosecution attorney'' model argues the prompt is a prompt injection, and a ``judge'' model gives the final classification. CourtGuard has a lower false positive rate than the Direct Detector, an LLM as-a-judge. However, CourtGuard is generally a worse prompt injection detector. Nevertheless, this lower false positive rate highlights the importance of considering both adversarial and benign scenarios for the classification of a prompt. Additionally, the relative performance of CourtGuard in comparison to other prompt injection classifiers advances the use of multiagent systems as a defense against prompt injection attacks. The implementations of CourtGuard and the Direct Detector with full prompts for Gemma-3-12b-it, Llama-3.3-8B, and Phi-4-mini-instruct are available at \url{https://github.com/isaacwu2000/CourtGuard}.
\end{abstract}

\section{Introduction}
LLM systems vulnerable to prompt injection attacks are increasingly able to access 
medical data, child conversation data, government and military intelligence, sensitive scientific research, financial data, and intellectual property (e.g. system prompts).
\citep{Zhan2025,Lee2025}. Beyond data leakage, prompt injection attacks can also alter the behavior of LLM systems. These hijacked systems can be used for spear phishing and disinformation through malicious websites or poisoned retrieval-augmented generation (RAG) applications \citep{Hazell2023,ClopTeglia2024,Rossi2024}.

However, despite a growing scientific literature and various enterprise solutions aiming to prevent prompt injections, it remains an open problem. For instance, even on Lakera AI’s own benchmark, the Prompt Injection Test (PINT), Lakera Guard scores only 92.5461\%, which implies that 7.5439\% of inputs were falsely classified \citep{LakeraND}. In addition, Nick Winter, VP of product and growth at Gray Swan AI, one of the premier AI security companies, says that “typically [prompt injection] attack is winning against defense” (N. Winter, personal communication, July 13, 2025).

Furthermore, while existing foundation model companies have created defenses such as Anthropic’s Constitutional Classifier, they are not fully robust, and many production applications lack sufficient defenses \citep{Sharma2025, Lumelsky2025}.

\section{Current Landscape}
Most current prompt injection defenses against can be divided into the following categories:

\begin{tabular}{p{3cm} p{6cm} p{6cm}}
\toprule
\textbf{Name} & \textbf{Description} & \textbf{Strengths and Weaknesses} \\
\midrule
Detection & 
Prompt injection detection uses techniques such as perplexity filtering \citep{AlonKamfonas2023}, where unnatural tokens are flagged, and fine-tuned LLMs to detect attacks inputted by the user or hidden in unsanitized text \citep{Sharma2025}. Additionally, fine-tuned LLMs can also do output filtering to detect when an LLM violates its system instructions \citep{Sharma2025}. & 
In some cases, input classifiers can be run concurrently with the main LLM system, and detection systems are generally effective. However, they have fundamental limitations in maximum accuracy and the model size of the detector \citep{Rao2024}. \\
\midrule
Separation & 
The splitting of system instructions from other tokens involves restricted delimiters \citep{Chen2024StruQ} or special tokens \citep{Suo2024} to grant certain permissions. & 
System prompt separation is effective and relatively easy to implement, but remains relatively vulnerable to prompt injections \citep{Chen2024StruQ}. \\
\midrule
Prompting & 
Prompt engineering system instructions can be done with self reminders to be responsible \citep{Xie2023} or the establishment of principles for the LLM to follow in the system prompt \citep{AnthropicND}. & 
While prompting is very easy to implement, it often performs poorly compared to other defenses. \citep{Watts2025}. \\
\midrule
Data handling & 
Data handling entails using system architectures such as CaMeL to prevent the LLM in contact with unsanitized tokens from accessing sensitive data, or simply keeping a human in-the-loop to grant permissions \citep{Debenedetti2025}. & 
While this approach inherently defends against prompt injections, it also limits the use cases of LLMs as they are barred from directly handling external data. In addition, these applications can cost over double that of those without data handling \citep{Debenedetti2025}. \\
\midrule
Combination & 
A combination of defenses can be used to fully safeguard the LLM system. This approach is often used by enterprise solutions to increase robustness \citep{Watts2025}. & 
Using multiple defenses improves the security of LLM systems, but also increases their cost and latency \citep{Watts2025}. \\
\bottomrule
\end{tabular}

\section{CourtGuard}
The enterprises that deal with sensitive data and are most at risk to prompt injection attacks may need to handle data internally for security reasons. However, there is a lack of simple and effective prompt injection defenses that developers can easily create internally and deploy locally. To address this gap, we propose CourtGuard, a multiagent, LLM-based system that uses local LLMs to detect prompt injections.

\begin{figure}[H]
    \centering
    \caption{\textit{Flowchart of CourtGuard}}
    \includegraphics[width=0.5\linewidth]{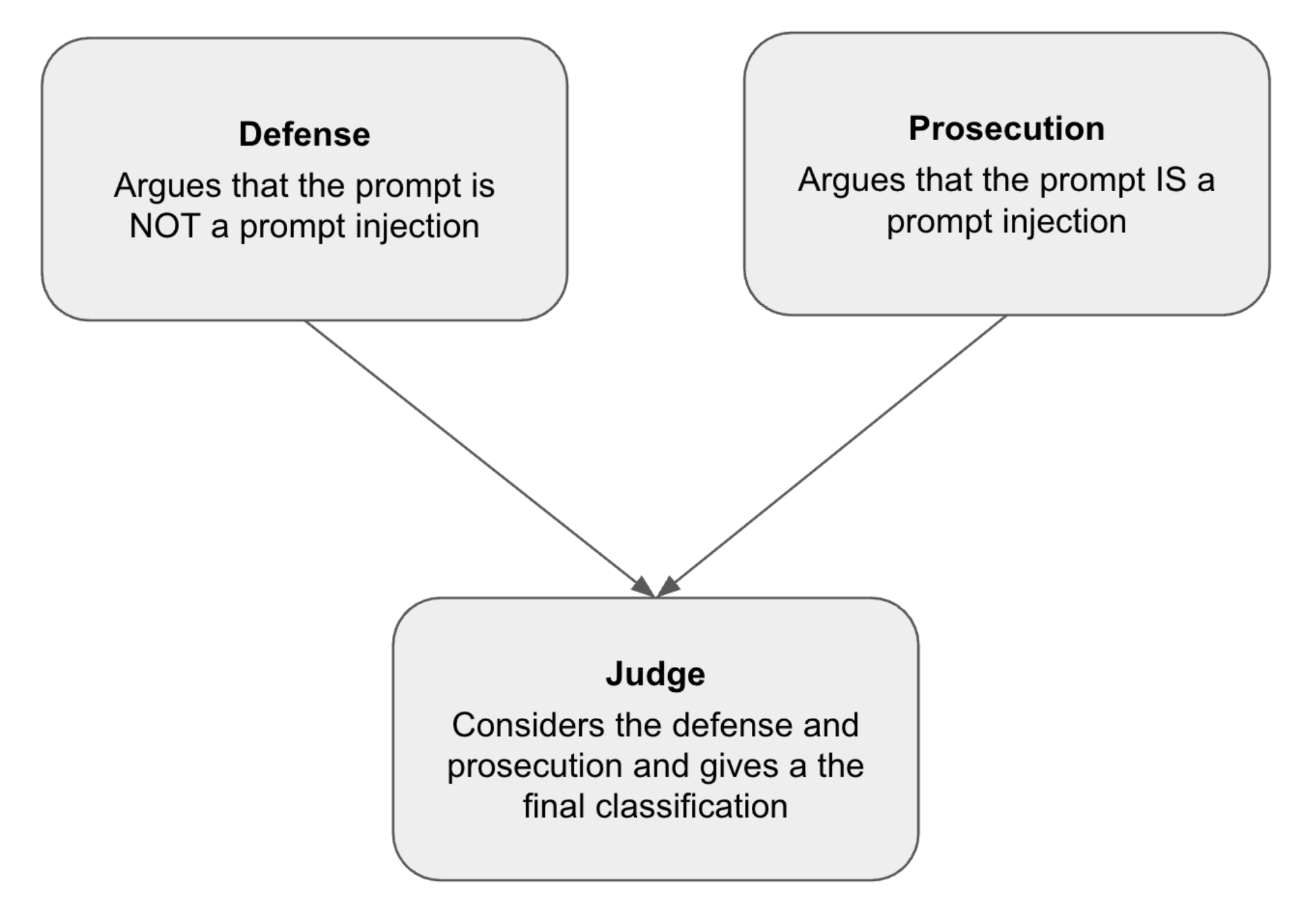}
    \label{fig:courtguard}
\end{figure}

First, two ``attorney models''—a ``defense'' and ``prosecution'' model—construct arguments in parallel for why the provided input is and is not a prompt injection respectively. Then, a ``judge model'' considers both arguments and issues a final verdict on whether the prompt is a prompt injection.

CourtGuard is a prompt injection detector, so no excess tokens are added to the system prompt of the LLMs in the main application. Additionally, given that the main application does not alter anything (e.g., write to a database), CourtGuard can be run in parallel with the LLM system. Therefore, in contrast to the multiagent system used for policy enforcement presented by \citet{Gosmar2025}, which runs after the main LLM system, an application using CourtGuard is more efficient.

\section{Methodology}

\subsection{Evaluation}

To evaluate CourtGuard’s ability to correctly classify both real-world prompt injections and benign prompts, we chose the following datasets for evaluation.

\begin{table}[h]
  \centering
  \begin{tabular}{p{0.25\linewidth} p{0.65\linewidth}}
    \toprule
    \textbf{Name} & \textbf{Description} \\
    \midrule
    LLMail-Inject & Contains a large variety of real-world, indirect attacks collected from participants in an email-based prompt injection challenge \citep{Abdelnabi2025a}. \\
    \addlinespace
    NotInject & Includes various benign prompts containing trigger words associated with prompt injections \citep{LiLiu2024}. \\
    \addlinespace
    Qualifire Prompt Injection Benchmark & A mix of both benign and jailbreak prompts to ensure accurate prompt injection detection with limited false positives \citep{QualifireND}. \\
    \bottomrule
  \end{tabular}
\end{table}

Since LLMail-Inject consists of over 460{,}000 prompt injection attacks \citep{Abdelnabi2025b}, a random sample, stratified by phase, of 5{,}000 attacks was selected as the evaluation dataset. Furthermore, since the prompts in the LLMail-Inject dataset had a separate subject and body, they were combined to form a single prompt using the method described in the appendix section titled ``Dataset Formatting.''

\subsection{Model Selection}

Because CourtGuard aims to improve local prompt injection defense, we chose to use Gemma-3-12b-it, Llama-3.3-8B, and Phi-4-mini-instruct because they are small enough to be run on a local computer; created by frontier foundation model companies, which suggests they have more support and production use; and powerful enough to (1) not be prompt injected themselves, (2) hallucinate minimally, and (3) effectively adhere to instructions. Additionally, when running, the temperature of all models was set to zero for reproducibility and optimal classification.

\subsection{Prompt Refinement}

A stratified random sample (three from each split) of nine prompts from the NotInject dataset was used for prompt optimization, so only the remaining 330 prompts were used for evaluation. Additionally, another small random sample of ten other attacks was selected from the LLMail-Inject dataset for prompt optimization.

Through this split, we could quickly optimize the prompts for both the Direct Detector and CourtGuard, as the low sample size allowed us to analyze individual outputs for each example. We optimized the prompts for the Direct Detector and CourtGuard system for Gemma by making small tweaks and observing whether they increased or decreased the scores on either sample. For consistency, the same prompts were used for all three models.

\section{Quantitative Analysis}

\subsection{Empirical Results}

\begin{table}[h]
  \centering
  \caption{Evaluation performances of the Direct Detector and CourtGuard.}
  \label{tab:eval-perf}
  \begin{tabular}{l l S[table-format=2.2] S[table-format=2.2] S[table-format=2.2]}
    \toprule
    \textbf{Model} & \textbf{System} & \textbf{LLMail-Inject (\%)} & \textbf{NotInject (\%)} & \textbf{Qualifire (\%)} \\
    \midrule
    gemma3-12b-it & Direct & 95.68 & 58.48 & 64.47\textsuperscript{**} \\
    gemma3-12b-it & Court  & 73.81\textsuperscript{*} & 73.64 & 54.96 \\
    llama-3.3-8b  & Direct & 59.82 & 90.61 & 75.92 \\
    llama-3.3-8b  & Court  & 14.88 & 94.55 & 66.34 \\
    phi4-mini     & Direct & 17.42 & 99.09 & 76.00 \\
    phi4-mini     & Court  & 25.63\textsuperscript{*} & 99.09 & 70.96 \\
    \bottomrule
  \end{tabular}

  \textsuperscript{*}Two prompts failed to run and were not included in percentage calculations. \\
  \textsuperscript{**}28 prompts failed to run due to evaluation software issues and were not included.
\end{table}

\begin{table}[h]
  \centering
  \caption{\centering Difference between CourtGuard and the Direct Detector’s evaluation performance (CourtGuard $-$ Direct Detector).}
  \label{tab:diff-perf}
  \begin{tabular}{l S[table-format=3.2] S[table-format=2.2] S[table-format=2.2] S[table-format=2.2]}
    \toprule
    \textbf{Model} & \textbf{LLMail-Inject (\%)} & \textbf{NotInject (\%)} & \textbf{Qualifire Benign (\%)} & \textbf{Qualifire Jailbreak (\%)} \\
    \midrule
    gemma3-12b-it & -21.87 & 15.15 & 6.62  & -33.91 \\
    llama-3.3-8b  & -44.94 &  3.94 & 1.53  & -26.26 \\
    phi4-mini     &   8.21 &  0.00 & 2.20  & -15.91 \\
    \midrule
    Mean          & -19.53 &  6.36 & 3.45  & -25.36 \\
    \bottomrule
  \end{tabular}
\end{table}

\begin{table}[h]
  \centering
  \caption{\centering Evaluation performance of the Direct Detector and CourtGuard on benign and jailbreak prompts in the Qualifire dataset.}
  \label{tab:qualifire-perf}
  \begin{tabular}{l l S[table-format=2.2] S[table-format=2.2]}
    \toprule
    \textbf{Model} & \textbf{System} & \textbf{Qualifire Benign (\%)} & \textbf{Qualifire Jailbreak (\%)} \\
    \midrule
    gemma3-12b-it & Direct & 41.86\textsuperscript{*} & 98.59\textsuperscript{**} \\
    gemma3-12b-it & Court  & 48.48 & 64.68 \\
    llama-3.3-8b  & Direct & 86.67 & 59.78 \\
    llama-3.3-8b  & Court  & 88.20 & 33.52 \\
    phi4-mini     & Direct & 94.57 & 48.12 \\
    phi4-mini     & Court  & 96.77 & 32.22 \\
    \bottomrule
  \end{tabular}
  
  \textsuperscript{*}Qualifire Benign = percentage of benign prompts correctly classified. \\
  \textsuperscript{**}Qualifire Jailbreak = percentage of jailbreaks correctly classified.
\end{table}

\subsubsection{Graphs of Evaluations of Direct Detector vs. CourtGuard for all Models}
\begin{figure}[H]
    \centering
    \caption{Direct Detector vs. CourtGuard Evals (Gemma)}
    \includegraphics[width=0.5\linewidth]{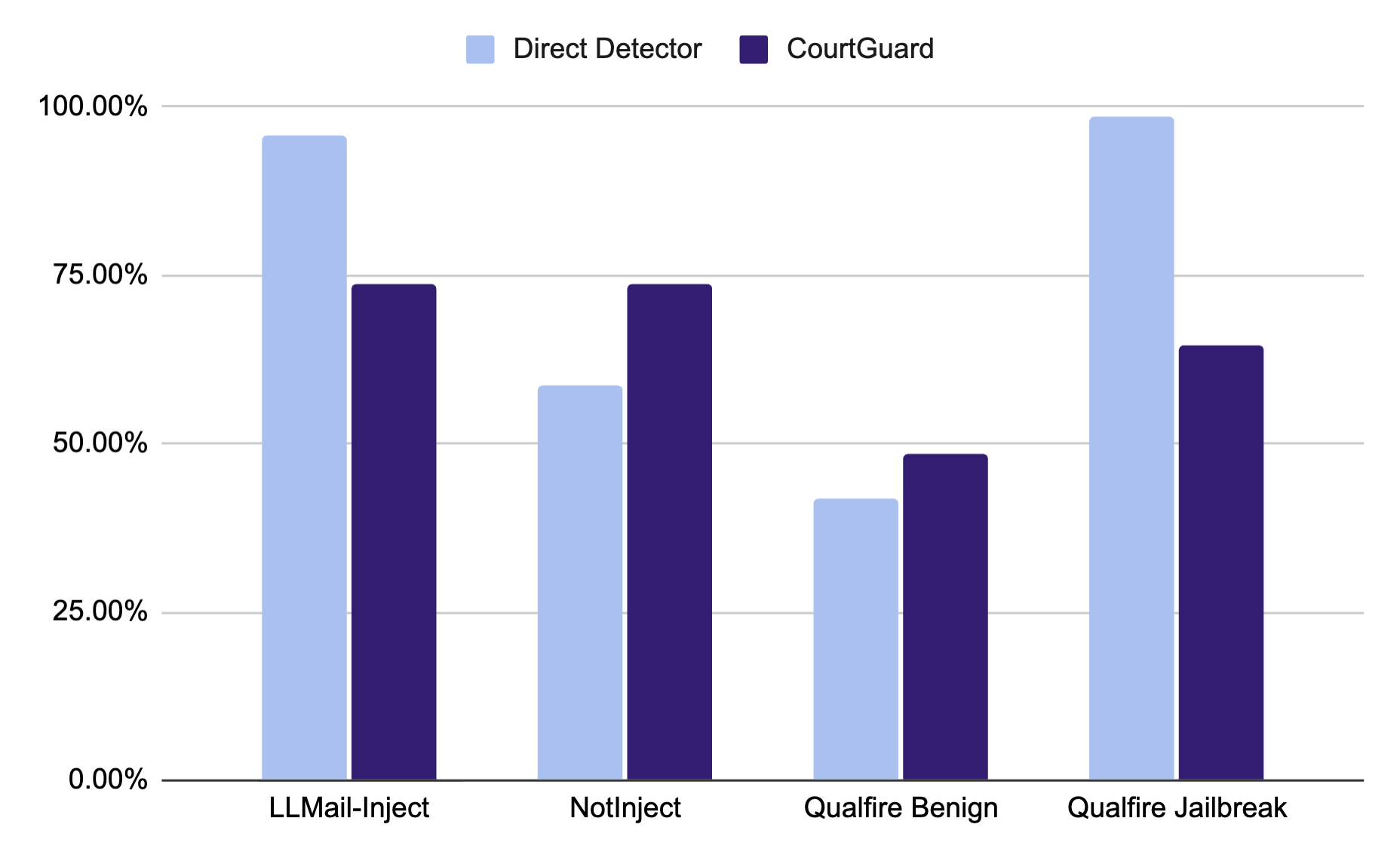}
    \label{fig:gemma_evals}
\end{figure}

\begin{figure}[H]
    \centering
    \caption{Direct Detector vs. CourtGuard Evals (Llama)}
    \includegraphics[width=0.5\linewidth]{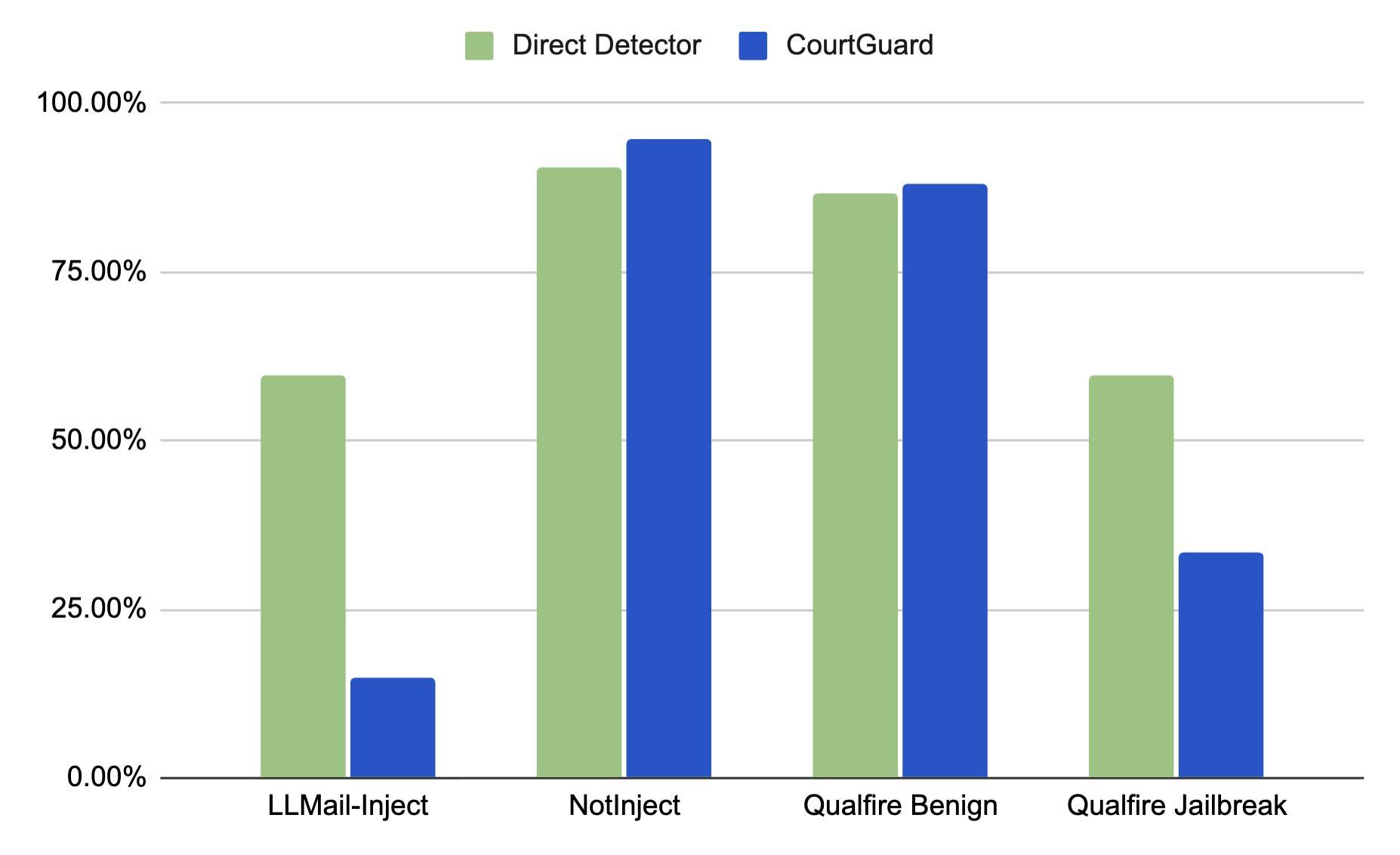}
    \label{fig:llama_evals}
\end{figure}

\begin{figure}[H]
    \centering
    \caption{Direct Detector vs. CourtGuard Evals (Phi)}
    \includegraphics[width=0.5\linewidth]{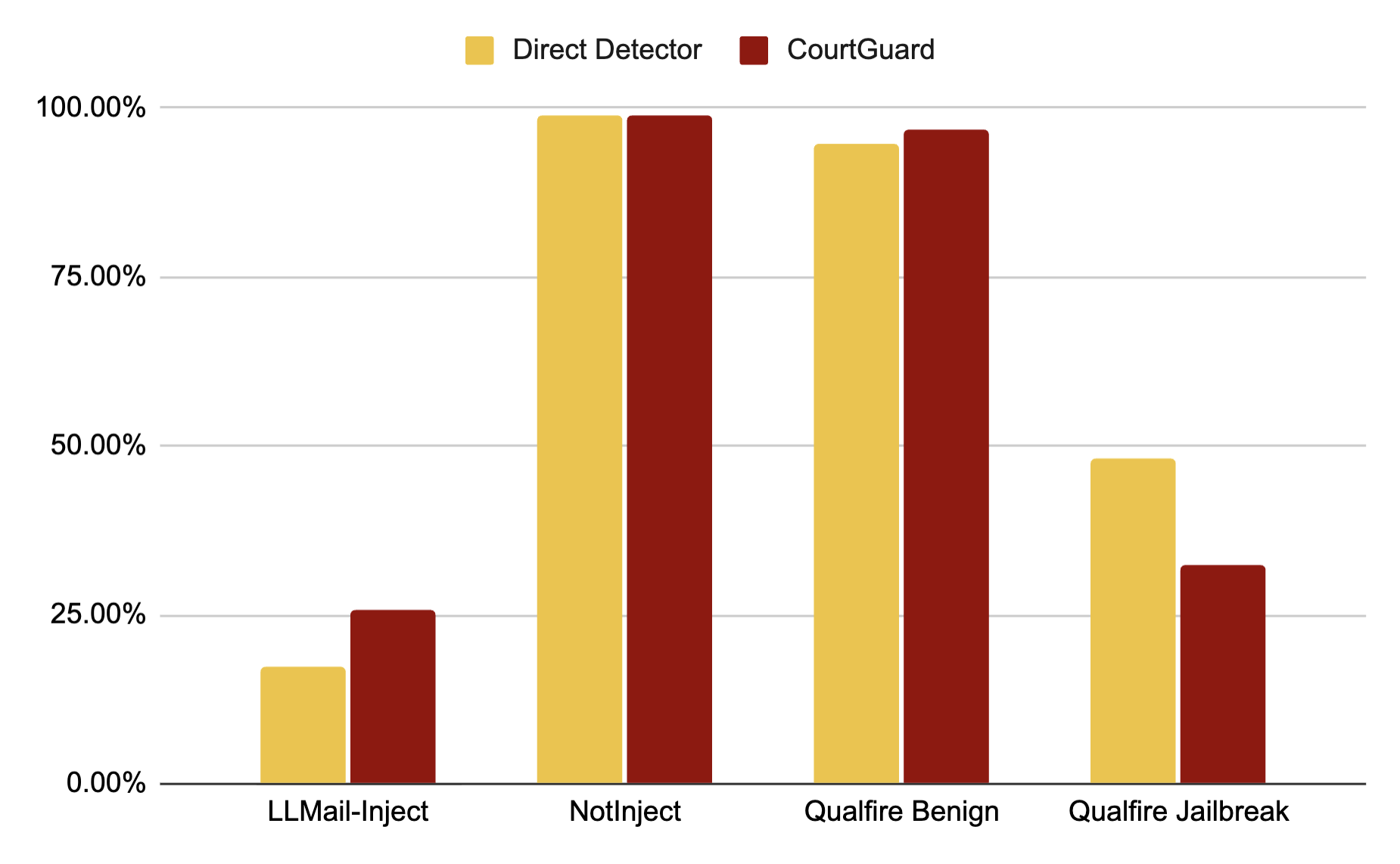}
    \label{fig:phi_evals}
\end{figure}

\subsubsection{Empirical Results Analysis}
In Table 2 and Figures 2 to 4 comparing the performance of the Direct Detector and CourtGuard on all evaluations, we see that, generally, CourtGuard correctly classified benign prompts at a higher percentage than the Direct Detector; CourtGuard scored higher on the NotInject dataset and correctly classified a higher percentage of benign prompts in the Qualifire Prompt Injection Benchmark.

Furthermore, the Direct Detector is generally better at classifying prompt injections, its classifying prompts from the LLMail-Inject dataset being roughly four times the percentage of CourtGuard (using Llama). However, there is one exception: Phi scores approximately 47\% higher on the LLMail-Inject dataset when using CourtGuard rather than the Direct Detector.

\subsection{Calculated Metrics}

\begin{table}[h]
  \centering
  \caption{Precision, recall, and F1 Score on the Qualifire Prompt Injection Benchmark.}
  \label{tab:metrics}
  \begin{tabular}{l l S[table-format=2.2] S[table-format=2.2] S[table-format=2.2]}
    \toprule
    \textbf{Model} & \textbf{System} & \textbf{Precision (\%)} & \textbf{Recall (\%)} & \textbf{F1 Score (\%)} \\
    \midrule
    gemma3-12b-it & Direct & 53.26 & 99.54 & 69.40 \\
    gemma3-12b-it & Court  & 59.26 & 83.10 & 69.18 \\
    llama-3.3-8b  & Direct & 75.02 & 59.78 & 66.54 \\
    llama-3.3-8b  & Court  & 67.00 & 33.96 & 45.07 \\
    phi4-mini     & Direct & 86.05 & 38.22 & 52.93 \\
    phi4-mini     & Court  & 87.38 & 32.39 & 47.27 \\
    \bottomrule
  \end{tabular}
\end{table}

\begin{table}[h]
  \centering
  \caption{Difference in precision, recall, and F1 score on the Qualifire dataset (CourtGuard $-$ Direct Detector).}
  \label{tab:diff-metrics}
  \begin{tabular}{l S[table-format=2.2] S[table-format=2.2] S[table-format=2.2]}
    \toprule
    \textbf{Model} & \textbf{Precision (\%)} & \textbf{Recall (\%)} & \textbf{F1 Score (\%)} \\
    \midrule
    gemma3-12b-it &  5.99 & -16.44 &  -0.21 \\
    llama-3.3-8b  & -8.02 & -25.82 & -21.46 \\
    phi4-mini     &  1.33 &  -5.83 &  -5.66 \\
    \midrule
    Mean          & -0.23 & -16.03 &  -9.11 \\
    \bottomrule
  \end{tabular}
\end{table}

\subsubsection{Graphs of Precision, Recall, and F1 Score on the Qualifire Prompt Injection Benchmark for the Direct Detector vs. CourtGuard for all Models}

\begin{figure}[H]
    \centering
    \caption{Direct Detector vs. CourtGuard Metrics (Gemma)}
    \includegraphics[width=0.5\linewidth]{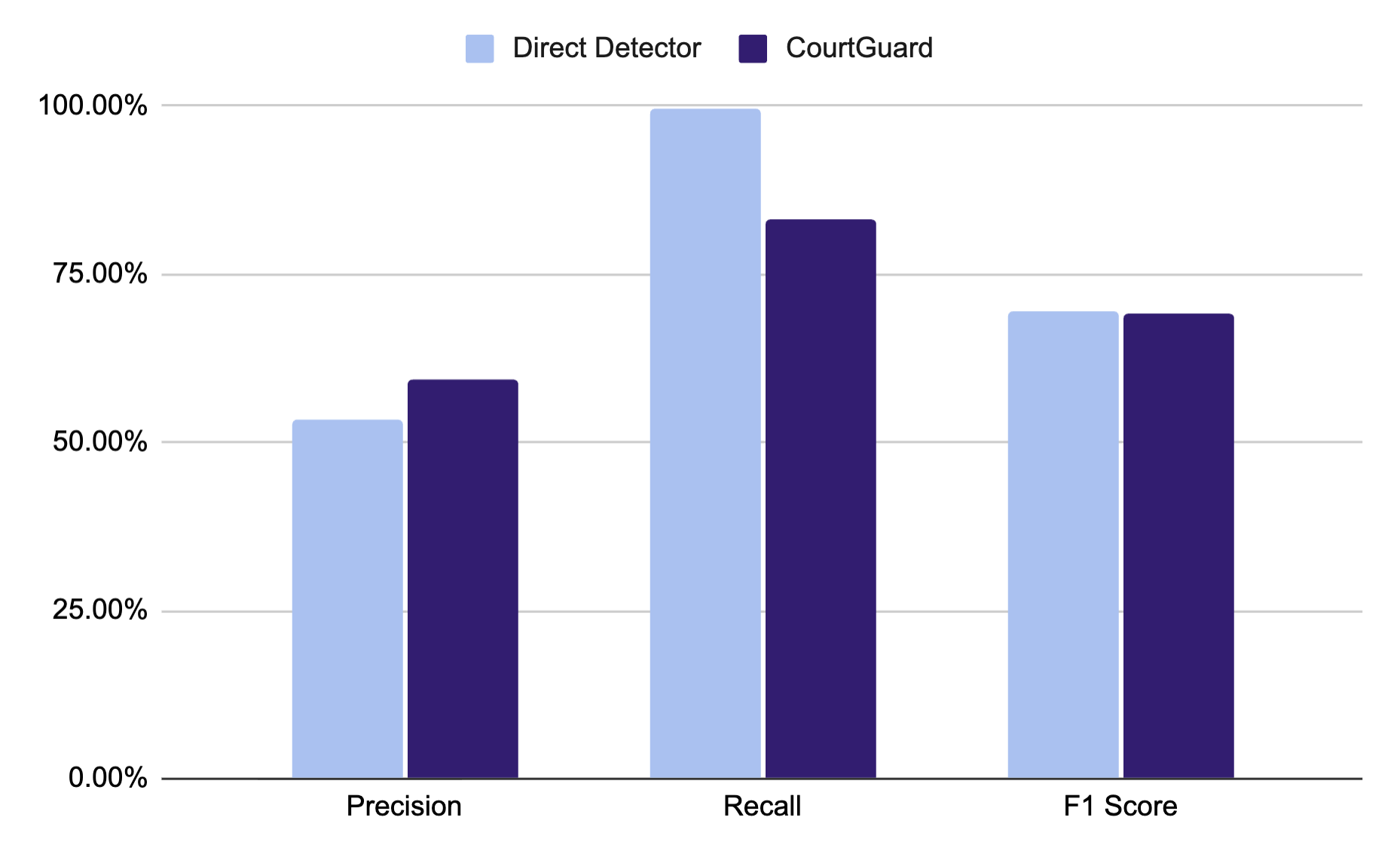}
    \label{fig:gemma_scores}
\end{figure}

\begin{figure}[H]
    \centering
    \caption{Direct Detector vs. CourtGuard Metrics (Llama)}
    \includegraphics[width=0.5\linewidth]{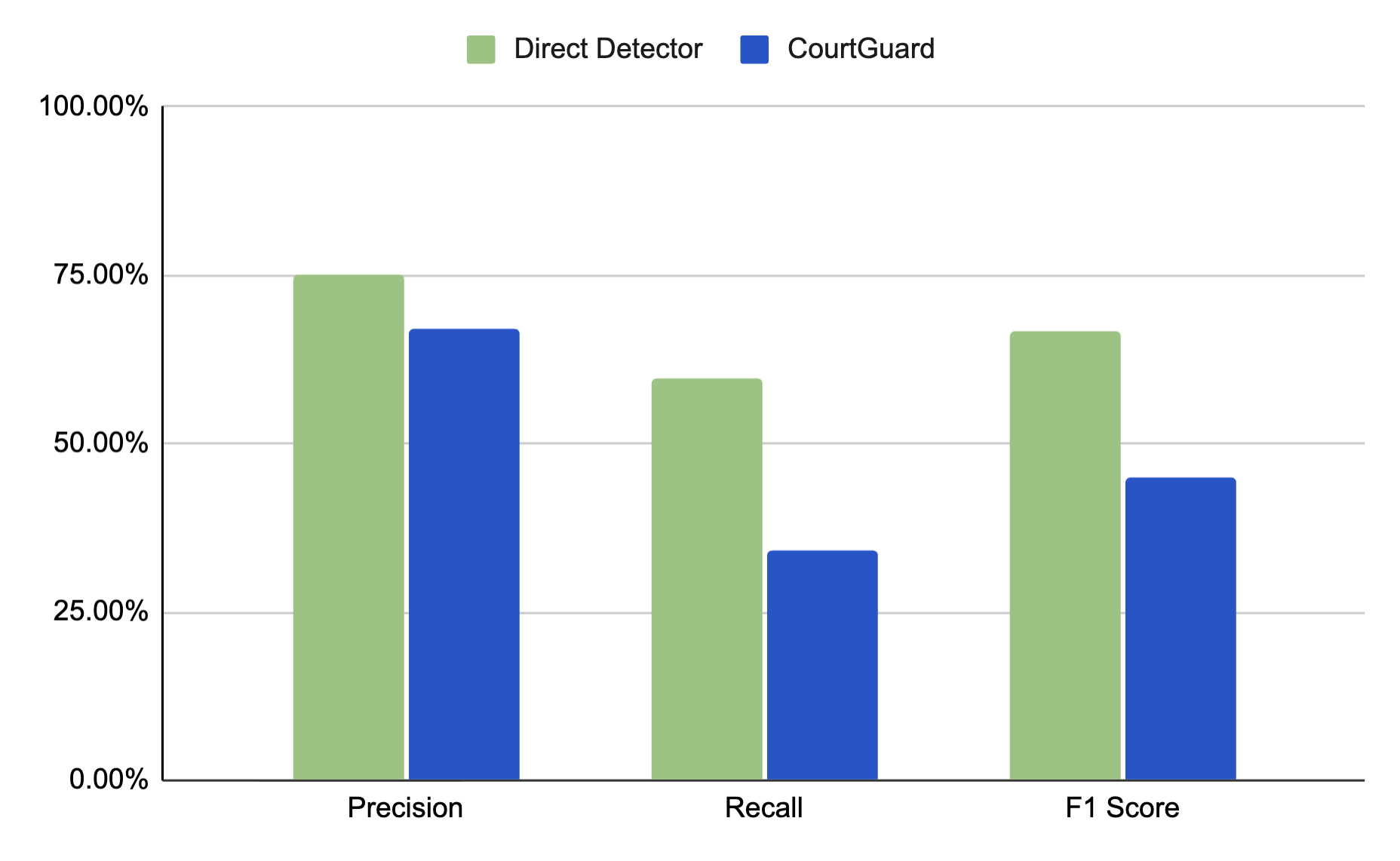}
    \label{fig:llama_scores}
\end{figure}

\begin{figure}[H]
    \centering
    \caption{Direct Detector vs. CourtGuard Metrics (Phi)}
    \includegraphics[width=0.5\linewidth]{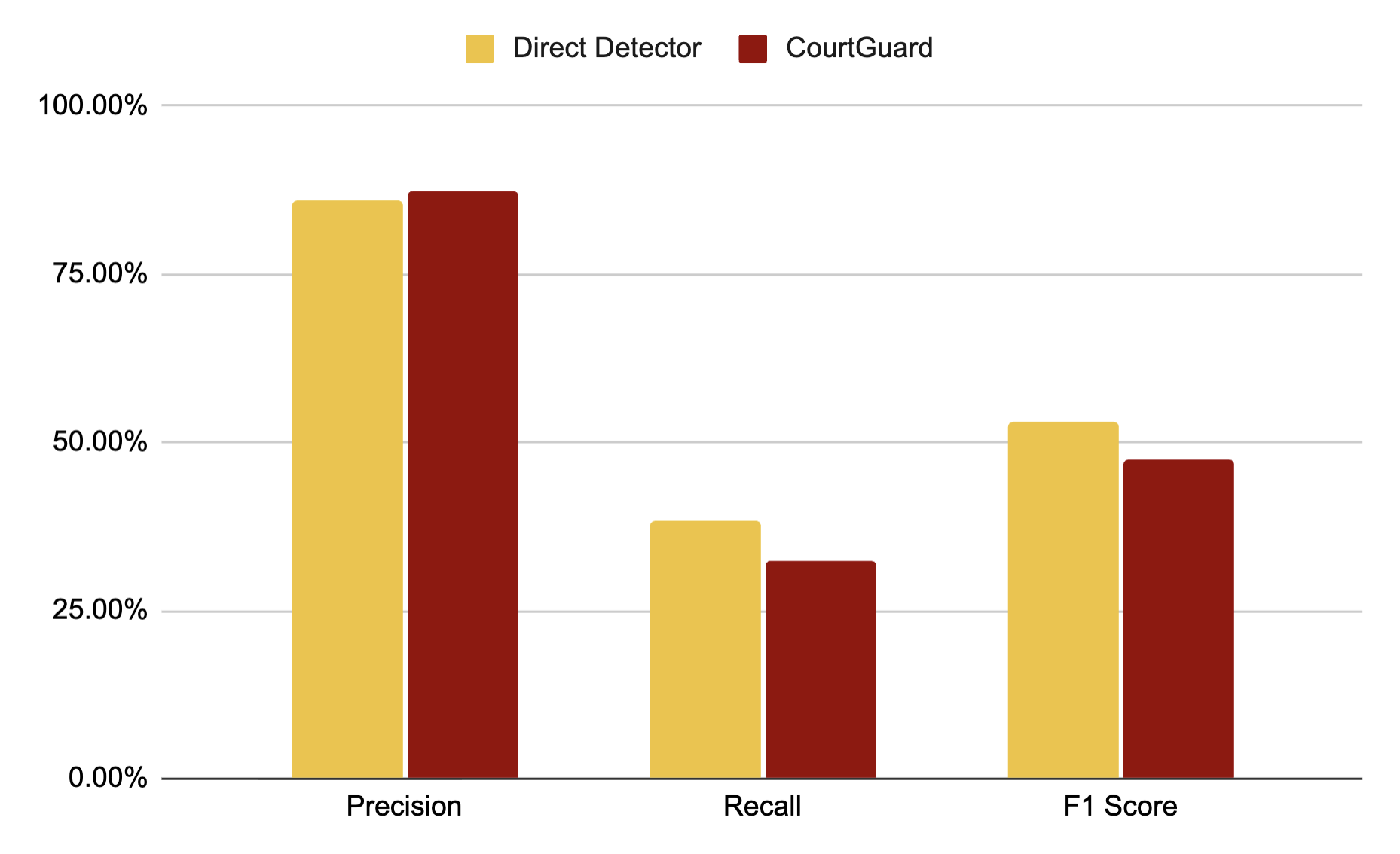}
    \label{fig:phi_scores}
\end{figure}

\subsubsection{Calculated Metrics Analysis}
In Table 4 as well as Figures 5 to 7, we see that in precision, recall, and F1 Score, the Direct Detector is generally higher or around the same as CourtGuard. These differences suggest that the Direct Detector is overall a better classifier of prompt injections in terms of balancing correct classifications and false positive rates, a balance which the F1 score measures.

\subsection{Comparison with Other Prompt Injection Solutions}
\subsubsection{LLMail-Inject}
Since this paper uses an updated version of the LLMail-Inject dataset—it includes Phase 2—no direct comparisons can be drawn between the Direct Detector and CourtGuard and other solutions.

\subsubsection{NotInject}
When using Llama or Phi, both the Direct Detector and CourtGuard perform fairly well on NotInject: all four combinations of defenses with the two models and two solutions score above 90\% on the NotInject benchmark, a score that exceeds Meta’s PromptGuard, LakeraGuard, GPT-4o, and even InjecGuard, a solution designed to prevent over defensiveness \citep{LiLiu2024}. Only Meta’s LlamaGuard3, which scored 99.71\% on the NotInject benchmark, exceeded the performances of all instances of the Direct Detector and CourtGuard \citep{LiLiu2024}. Even so, the Direct Detector and CourtGuard using Phi both scored a similar 99.09\%.

\subsubsection{Qualifire Prompt Injection Benchmark}
Qualifire’s Sentinel model, which was self proclaimed as state-of-the-art in its release in June, 2025, scores 97.6\% on the Qualifire Prompt Injection Benchmark—much higher than all instances of the Direct Detector or CourtGuard \citep{IvryNahum2025}. In contrast, ProtectAI’s fine-tuned deberta-v3-base-prompt-injection-v2 scores only 65.2\% on the benchmark, a score lower than that of the Direct Detector and CourtGuard using Llama or Phi \citep{IvryNahum2025}.

However, both of these solutions are enterprise solutions, Sentinel being only 0.395B parameters and deberta-v3-base-prompt-injection-v2 being only 0.185 parameters. Thus, Sentinel having an average latency of only around 0.02 seconds per inference time, they both have a significantly lower inference time compared to the Direct Detector and CourtGuard. 

\section{Qualitative Analysis}
The Direct Detector often appears to directly assume, perhaps to sound more coherent, whether the prompt is a prompt injection before it creates its response. Occasionally, despite being prompted not to, the Direct Detector, for all models, will state its classification at the start of its response. For example, the Direct Detector (Phi) in one case already writes in the second sentence, without any reasoning beforehand, that “there is no indication of an attempt to manipulate or exploit a system,” an immediate classification of the prompt without any consideration of alternatives. In another, more subtle example, the Direct Detector (Gemma) starts its response with, ``Okay, let's analyze the user input: [input redacted for safety] \textbackslash n\textbackslash n**Analysis:**\textbackslash n\textbackslash n*   **Surface Appearance:** At first glance, this appears to be a straightforward, innocent question.'' This start already implies that the prompt is a prompt injection as shown by the use of the phrase ``at first glance.'' 

In contrast, the judge in CourtGuard, by the nature of the multiagent framework and prompt, almost always appears to consider both possible classifications:

\begin{quote}
After analyzing both the defense and prosecution arguments, I will provide a comprehensive evaluation of the case.\textbackslash n\textbackslash nThe defense argument…On the other hand, the prosecution argument…Upon closer examination of the original prompt…After careful consideration of both arguments, I conclude…**Final Verdict:** I am moderately confident that the prompt, on its own, is a prompt injection.
\end{quote}

From this structure, it is apparent that CourtGuard considers both possibilities and then makes a final conclusion (though it is possible that the conclusion is somehow encoded in previous tokens).

Thus, perhaps similarly to the hidden reasoning described by \citet{Chen2025ReasoningModels}, the assuming approach used by the Direct Detector could have allowed it to more easily detect prompt injections by directly asserting a prompt was a prompt injection based non-verbalized encodings. In contrast, CourtGuard has to consider both the adversarial and benign scenario and thus has a lower false positive rate when classifying prompts. However, CourtGuard's approach also relies on the model's reasoning abilities, which results in lower true positive scores.

\section{Limitations}
A major limitation of LLM prompt injection classifiers is their inference time, which can reach several seconds when running a LLM locally. In the future, researchers could optimize latency by parallelizing inference, exploring smaller, open-weight LLMs, and quantization. Additionally, CourtGuard is not robust enough to fully protect LLM systems, so further enhancements are necessary. Furthermore, CourtGuard and the Direct Detector were only tested on public datasets of static, singular prompt injection attacks. Real-world prompt injection attacks are often implemented throughout multi-turn conversations and with adaptive attacks, both of which were not tested in this paper. Evaluating on these real-world prompt injection attacks would be a critical addition for practicality and as a control against data contamination, especially if a system like CourtGuard were to be employed in production.

\section{Conclusion}
Overall, CourtGuard is better than the Direct Detector at correctly classifying benign prompts, while the Direct Detector is better at correctly classifying prompt injections. Furthermore, the Direct Detector’s F1 Score on the Qualifire Prompt Injection Benchmark being higher than CourtGuard’s for all models suggests that the Direct Detector is a superior general prompt injection classifier. These results are likely due to the Direct Detector's assuming the classification before beginning its response, an approach that perhaps allows the model to rely on hidden thoughts based on potential training on prompt injection classification. Contrarily, the Judge model in CourtGuard considers both possible classifications of a prompt in a novel manner, which could force the model to fully articulate its reasoning. However, without the hidden thoughts to leverage the model’s potential previous training on prompt injection, CourtGuard has a lower ability to classify true prompt injections.

Therefore, future researchers should ensure that their prompt injection classifiers consider both possible classifications for a prompt to reduce false positives. In addition, multiagent systems and local LLMs show promise for prompt injection defense, especially when latency is a lower priority, as CourtGuard and the Direct Detector both score decently on the NotInject and Qualifire datasets compared to other solutions. Thus, as simple and locally runnable approaches like CourtGuard develop, AI developers in data-sensitive enterprises should consider employing them for LLM applications in contact with sensitive data.

\section{Acknowledgments}
We thank the Non-Trivial Ventures for providing the structured fellowship which made this paper possible and Nick Winter, Wenxiao Wang, and Alex Wilf for their expert advice during ideation.

\appendix
\section{Dataset Formatting}
Prompts in the LLMail-Inject dataset had a subject and body field, which we unified by adding a semicolon and a space after the subject during concatenation.

For example, the dictionary
\begin{verbatim}
{"subject": <subject>, "body": <body>}
\end{verbatim}
would become the string
\begin{verbatim}
f"subject: {subject}; body: {body}"
\end{verbatim}



\begin{thebibliography}{}

\bibitem[Abdelnabi et~al.(2025a)Abdelnabi et~al.]{Abdelnabi2025a}
Abdelnabi, S., Fay, A., Salem, A., Zverev, E., Liao, K.-C., Liu, C.-H., Kuo, C.-C., Weigend, J., Manlangit, D., Apostolov, A., Umair, H., Donato, J., Kawakita, M., Mahboob, A., Bach, T. H., Chiang, T.-H., Cho, M., Choi, H., Kim, B., Lee, H., Pannell, B., McCauley, C., Russinovich, M., Paverd, A., \& Cherubin, G. (2025).
LLMail-inject: A dataset from a realistic adaptive prompt injection challenge.
\textit{arXiv preprint}. \url{https://doi.org/10.48550/arXiv.2506.09956}

\bibitem[Abdelnabi et~al.(2025b)Abdelnabi et~al.]{Abdelnabi2025b}
Abdelnabi, S., Fay, A., Salem, A., Zverev, E., Liu, C.-H., Kuo, C.-C., Weigend, J., Manlangit, D., Apostolov, A., Umair, H., Donato, J., Kawakita, M., Mahboob, A., Bach, T. H., Chiang, T.-H., Cho, M., Choi, H., Kim, B., Lee, H., Pannell, B., McCauley, C., Russinovich, M., Paverd, A., \& Cherubin, G. (2025).
\textit{LLMail-inject challenge} [Data set].
Hugging Face. \url{https://huggingface.co/datasets/microsoft/llmail-inject-challenge}

\bibitem[Alon \& Kamfonas(2023)]{AlonKamfonas2023}
Alon, G., \& Kamfonas, M. (2023).
Detecting language model attacks with perplexity.
\textit{arXiv preprint}. \url{https://doi.org/10.48550/arXiv.2308.14132}

\bibitem[Anthropic(n.d.)]{AnthropicND}
Anthropic. (n.d.).
Mitigate jailbreaks and prompt injections.
Retrieved September 11, 2025, from \url{https://docs.anthropic.com/en/docs/test-and-evaluate/strengthen-guardrails/mitigate-jailbreaks}

\bibitem[Chen et~al.(2025)Chen, Benton, Radhakrishnan, Uesato, Denison, Schulman, Somani, Hase, Wagner, Roger, Mikulik, Bowman, Leike, Kaplan, Perez]{Chen2025ReasoningModels}
Chen, Y., Benton, J., Radhakrishnan, A., Uesato, J., Denison, C., Schulman, J., Somani, A., Hase, P., Wagner, M., Roger, F., Mikulik, V., Bowman, S., Leike, J., Kaplan, J., \& Perez, E. (2025).  
**Reasoning Models Don’t Always Say What They Think**.  
Anthropic. \url{https://assets.anthropic.com/m/71876fabef0f0ed4/original/reasoning_models_paper.pdf}

\bibitem[Chen et~al.(2024)Chen, Piet, Sitawarin, \& Wagner]{Chen2024StruQ}
Chen, S., Piet, J., Sitawarin, C., \& Wagner, D. (2024).
StruQ: Defending against prompt injection with structured queries.
\textit{arXiv preprint}. \url{https://doi.org/10.48550/arXiv.2402.06363}

\bibitem[Clop \& Teglia(2024)]{ClopTeglia2024}
Clop, C., \& Teglia, Y. (2024).
Backdoored retrievers for prompt injection attacks on retrieval augmented generation of large language models. \textit{arXiv preprint}. \url{https://doi.org/10.48550/arXiv.2410.14479}

\bibitem[Debenedetti et~al.(2025)]{Debenedetti2025}
Debenedetti, E., Shumailov, I., Fan, T., Hayes, J., Carlini, N., Fabian, D., Kern, C., Shi, C., Terzis, A., \& Tramèr, F. (2025).
Defeating prompt injections by design.
\textit{arXiv preprint}. \url{https://doi.org/10.48550/arXiv.2503.18813}

\bibitem[Gosmar et~al.(2025)]{Gosmar2025}
Gosmar, D., Dahl, D. A., \& Gosmar, D. (2025).
Prompt injection detection and mitigation via AI multi-agent NLP frameworks.
\textit{arXiv preprint}. \url{https://doi.org/10.48550/arXiv.2503.11517}

\bibitem[Hazell(2023)]{Hazell2023}
Hazell, J. (2023).
Spear phishing with large language models.
\textit{arXiv preprint}. \url{https://doi.org/10.48550/arXiv.2305.06972}

\bibitem[Ivry \& Nahum(2025)]{IvryNahum2025}
Ivry, D., \& Nahum, O. (2025).
Sentinel: SOTA model to protect against prompt injections.
\textit{arXiv preprint}. \url{https://doi.org/10.48550/arXiv.2506.05446}

\bibitem[Lakera AI(n.d.)]{LakeraND}
Lakera AI. (n.d.).
Lakera PINT benchmark. 
GitHub. Retrieved September 11, 2025, from \url{https://github.com/lakeraai/pint-benchmark}

\bibitem[Lee et~al.(2025)]{Lee2025}
Lee, Y., Park, T., Lee, Y., Gong, J., \& Kang, J. (2025).
Exploring potential prompt injection attacks in federated military LLMs and their mitigation.
\textit{arXiv preprint}. \url{https://doi.org/10.48550/arXiv.2501.18416}

\bibitem[Li \& Liu(2024)]{LiLiu2024}
Li, H., \& Liu, X. (2024).
InjecGuard: Benchmarking and mitigating over-defense in prompt injection guardrail models.
\textit{arXiv preprint}. \url{https://doi.org/10.48550/arXiv.2410.22770}

\bibitem[Lumelsky(2025)]{Lumelsky2025}
Lumelsky, A. (2025).
OWASP top 10 LLM, updated 2025: Examples and mitigation strategies.
Oligo Security Academy. Retrieved September 11, 2025, from \url{https://www.oligo.security/academy/owasp-top-10-llm-updated-2025-examples-and-mitigation-strategies}

\bibitem[Qualifire(n.d.)]{QualifireND}
Qualifire. (n.d.).
Prompt injections benchmark [Data set].
Hugging Face. Retrieved September 10, 2025, from \url{https://huggingface.co/datasets/qualifire/prompt-injections-benchmark}

\bibitem[Rao et~al.(2024)]{Rao2024}
Rao, A., Choudhury, M., \& Aditya, S. (2024).
[WIP] Jailbreak paradox: The Achilles' heel of LLMs.
\textit{arXiv preprint}. \url{https://doi.org/10.48550/arXiv.2406.12702}

\bibitem[Rossi et~al.(2024)Rossi, Michel, Mukkamala, \& Thatcher]{Rossi2024}
Rossi, S., Michel, A. M., Mukkamala, R. R., \& Thatcher, J. B. (2024).
An early categorization of prompt injection attacks on large language models.
\textit{arXiv preprint}. \url{https://doi.org/10.48550/arXiv.2402.00898}

\bibitem[Sharma et~al.(2025)]{Sharma2025}
Sharma, M., Tong, M., Mu, J., Wei, J., Kruthoff, J., Goodfriend, S., Ong, E., Peng, A., Agarwal, R., Anil, C., Askell, A., Bailey, N., Benton, J., Bluemke, E., Bowman, S. R., Christiansen, E., Cunningham, H., Dau, A., Gopal, A., Gilson, R., Graham, L., Howard, L., Kalra, N., Lee, T., Lin, K., Lofgren, P., Mosconi, F., O'Hara, C., Olsson, C., Petrini, L., Rajani, S., Saxena, N., Silverstein, A., Singh, T., Sumers, T., Tang, L., Troy, K. K., Weisser, C., Zhong, R., Zhou, G., Leike, J., Kaplan, J., \& Perez, E. (2025).
Constitutional classifiers: Defending against universal jailbreaks across thousands of hours of red teaming.
\textit{arXiv preprint}. \url{https://doi.org/10.48550/arXiv.2501.18837}

\bibitem[Suo(2024)]{Suo2024}
Suo, X. (2024).
Signed-prompt: A new approach to prevent prompt injection attacks against LLM-integrated applications.
\textit{arXiv preprint}. \url{https://doi.org/10.48550/arXiv.2401.07612}

\bibitem[Watts(2025)]{Watts2025}
Watts, S. (2025).
Prompt injection \& the rise of prompt attacks: All you need to know.
Lakera. Retrieved September 11, 2025, from \url{https://www.lakera.ai/blog/guide-to-prompt-injection}

\bibitem[Xie et~al.(2023)]{Xie2023}
Xie, Y., Yi, J., Shao, J., Curl, J., Lyu, L., Chen, Q., Xie, X., \& Wu, F. (2023).
Defending ChatGPT against jailbreak attack via self-reminders.
\textit{Nature Machine Intelligence, 5}(12), 1486–1496. \url{https://doi.org/10.1038/s42256-023-00765-8}

\bibitem[Zhan et~al.(2025)]{Zhan2025}
Zhan, Q., Fang, R., Panchal, H. S., \& Kang, D. (2025).
Adaptive attacks break defenses against indirect prompt injection attacks on LLM agents.
\textit{arXiv preprint}. \url{https://doi.org/10.48550/arXiv.2503.00061}

\end{thebibliography}
\end{document}